\documentclass[%
 reprint,
superscriptaddress,
 amsmath,amssymb,
pra,
]{revtex4-1}

\usepackage{graphicx}
\usepackage{dcolumn}
\usepackage{bm}

\begin{document}


\title{Density-Matrix Simulation of Logical Qubit \\using 3-qubit Quantum Error Correction Code}

\author{Chungheon Baek}
 \email{CHBaek@etri.re.kr}
\affiliation{Electronics and Telecommunications Research Institute (ETRI), Daejeon 34129, South Korea}
\author{Tomohiro Ostuka}%
\affiliation{Research Institute of Electrical Communication, Tohoku University, 2-1-1 Katahira, Aoba-ku, Sendai 980-8577, Japan
}
\affiliation{RIKEN, Center for Emergent Matter Science (CEMS), Wako-shi, Saitama 351-0198, Japan}
\affiliation{Department of Applied Physics, University of Tokyo, 7-3-1 Hongo, Bunkyo-ku, Tokyo 113-8656, Japan}
\affiliation{JST, PRESTO, 4-1-8 Honcho, Kawaguchi, Saitama 332-0012, Japan}

\author{Seigo Tarucha}%
\affiliation{RIKEN, Center for Emergent Matter Science (CEMS), Wako-shi, Saitama 351-0198, Japan}
\affiliation{Department of Applied Physics, University of Tokyo, 7-3-1 Hongo, Bunkyo-ku, Tokyo 113-8656, Japan}

\author{Byung-Soo Choi}%
 \email{bschoi3@etri.re.kr, corresponding author}
\affiliation{Electronics and Telecommunications Research Institute (ETRI), Daejeon 34129, South Korea}

\date{\today}

\begin{abstract}
Fault-tolerant quantum computing demands many qubits with long lifetimes to conduct accurate quantum gate operations. However, external noise limits the computing time of physical qubits. Quantum error correction codes may extend such limits, but imperfect gate operations introduce errors to the correction procedure as well. The additional gate operations required due to the physical layout of qubits exacerbate the situation. Here, we use density-matrix simulations to investigate the performance change of logical qubits according to quantum error correction codes and qubit layouts and the expected performance of logical qubits with gate operation time and gate error rates. 
Considering current qubit technology, the small quantum error correction codes are chosen.
Assuming 0.1\% gate error probability, a logical qubit encoded by a 5-qubit quantum error correction code is expected to have a fidelity 0.25 higher than its physical counterpart.

\end{abstract}

\maketitle

\section{\label{sec:level1}INTRODUCTION}
The increasing need for powerful computers has attracted much attention to quantum computers on the application to big data searching, quantum chemistry, machine learning, and quantum cryptography \cite{Shor1994,Grover1996,Buluta2009, Biamonte2017, Otterbach2017}. Quantum computers are expected to surpass the computational power of their classical counterparts when a system with approximately over 50 qubits has been realized. In anticipation of quantum  supremacy, scalable qubit systems have been under thorough investigation. Google \cite{Google49}, IBM \cite{IBM50}, and Intel \cite{Intel49} have announced that they aim to show such supremacy using 48 or more superconducting qubits. A blueprint for a 250-qubit system has been suggested for ion-trap qubits \cite{ion5:50,Lekitsch2015}, and single chip qubit array fabrication schemes have been reported for silicon quantum dot qubits\cite{Veldhorst2016,arXiv1711.03807}.

The ultimate goal is to conduct a fault-tolerant quantum computing for various quantum algorithms. Then, physical qubits are insufficient due to their limited reliability, and logical qubits become essential. In logical qubits, the original quantum information is distributed over multiple physical qubits. If they satisfy the maximum tolerable error threhold determined by quantum error correction(QEC) codes, errors can be detected then corrected \cite{Raussendorf2007,Aliferis2007,Steane2003,Gottesman2009}. The increased accuracy of logical qubits allows more reliable and frequent operation within the computing time without irreversible loss of data \cite{Jones2012}. Even though leading researchers deal with about 50 qubits for superconductors, promising candidates such as quantum dot and ion qubits have not yet implemented a large number of qubits. As an intermediate step, the QEC codes, even though it is not a fault-tolerant QEC code, composed of the small number of qubits are needed.

It is important to analyze how high the maximum tolerable error rate required for physical qubits. In a linear approximation model \cite{Sohn2017}, the error is reflected on discrete X and Z errors. It considers the relaxation and dephasing errors in the same way. In addition, the error propagation due to 2-qubit gate operation is ignored. Admitted that errors can be interpreted in a discrete and independent way, continuous minute error in infinitesimal time is close to the errors in physical qubits.

In this work, we aim to analyze more precise effect on the performance of logical qubits, using density-matrix simulation. This simulation is expected to describe quantum state close to the physical qubits, because it keeps track of the change in quantum state. We anticipate that the performance of the logical qubit as predicted by this simulation will be analyzed in more sophisticated way compared to the linear approximation model. We investigate the improvement of several specific states so that expected features of intermediate QEC codes are visualized. To support the density-matrix simulation result, it is shown to be consistent with the IBM quantum experience results conducted by physical qubits.

We suggest three application examples to use density-matrix simulation. First, the logical qubits encoded by different QEC codes can be analyzed. Even though density-matrix simulation can analyze various QEC codes, we focused on 3-qubit QEC codes, which needs a small number of qubits without syndrome measurements, to verify 5-qubit QEC code can correct both X and Z errors. Second, it is possible to estimate how much the minimum error rate for implementing a logical qubit should be in a situation where nearest neighbor operations are required. QEC codes which needs large number of qubits may keep the quantum state in a longer time, but require to operate more complicated operations. In terms of cost and benefit, one can compare the estimated performance of logical qubits in different layouts. Finally, the current status of physical qubits can be evaluated. Assuming that a sufficient number of qubits with the current data of physical qubits, the estimated performance of logical qubits can be estimated. In the simulation, considering the gate error of 0.1\% and the gate operation time of a thousandth of coherence time in an all-to-all connected layout, the fidelity of logical qubits encoded by 5-qubit QEC codes is expected to be higher than that of physical qubit by 0.25.

\section{\label{sec:level1}PRELIMINARIES}
\subsection{\label{sec:level2}3-qubit quantum error correction code}
\begin{figure}
\includegraphics[width=\linewidth]{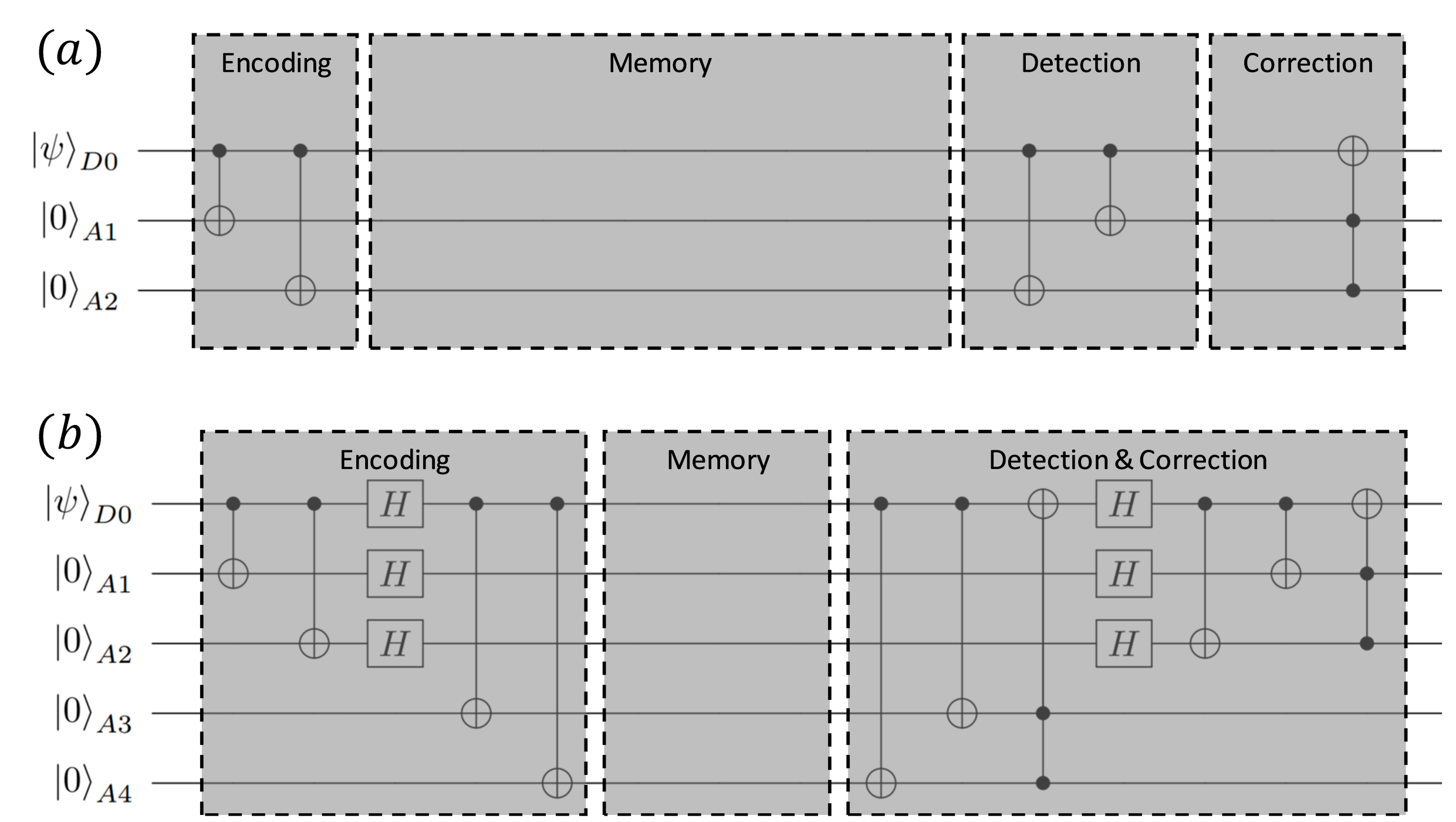}
\caption{\label{fig:circuit} (a) 3-qubit bit-flip QEC circuit. QEC procedures are composed of 4 parts: encoding, memory time, detection, and correction. Memory time is the computing time of logical qubit. QEC time is time taken by the other 3 procedures. (b) 5-qubit QEC circuit. bit and phase flip error are corrected by the first and second toffoli gates, respectively}
\end{figure}

3-qubit QEC codes have been experimentally conducted in many physical systems\cite{Reed2012,Kelly2015,Riste2015,Corcoles2015,Ofek2016}.
3-qubit bit-flip QEC circuit consists of the following four parts, shown in Fig. \ref{fig:circuit}(a). The first qubit is data qubit, and two ancilla qubits are prepared in $|0\rangle$ state. In the encoding process, two CNOT gates make logical $|0\rangle$ and $|1\rangle$ states as $|000\rangle$ and $|111\rangle$, respectively. In memory time, logical qubit gate operations can be conducted, just as a physical qubit gate operation can be done within lifetime. Therefore, the memory time can be interpreted as logical qubit's computing time, as the counterpart of lifetime of physical qubit. In the detection and correction procedure, a bit flip error occurring in data qubit is corrected. QEC time is defined as the sum of encoding, detection and correction times. Finally, the restored quantum state is detected in a measurement process.

To analyze the positive gain of fidelity using QEC, error sources are divided into dephasing and gate error.
Over the entire procedures, all qubits independently suffer from dephasing error, relying on the $T_1$ and $T_2$ coherence time. For generality, the coherence times is set as 1. Quantum gate also undergoes errors. In order to increase computing time, the comparison between additional dephasing error in memory time and correcting power by QEC code is compared. The errors can be reduced in a practical manner. Suppressing environmental noise may suppress the dephasing errors \cite{Yoneda2018,Muhonen2014,Kuhlmann2013}. Gradient ascent pulse engineering may take short time to conduct the same gate operation \cite{Khaneja2005,DeFouquieres2011}.

5-qubit QEC code can correct X- and Z-type errors without syndrome measurement.
In spite of the simplicity of 3-qubit QEC code, it merely deals with a single type error. Even though the repetition of bit-flip and phase-flip codes, the remaining errors are accumulated rather than restored. On the other hand, 5-qubit QEC code can correct a single error on a data qubit. If a Y-type error is happened in data qubit, X error is dealt with the first toffoli gates and remaining Z error, changed into X error by hadamard gate, is done with the second toffoli gate. The circuit is composed of four parts: encoding, memory and detection and correction, which is illustrated in Fig. \ref{fig:circuit}(b).

\subsection{\label{sec:level2} Linear Approximation Model}

One way to evaluate the closeness of two quantum states is the fidelity. The value of fidelity between pure states such as $\rho=|\psi\rangle\langle\psi|$ and $\sigma =|\phi\rangle \langle\phi|$ can be interpreted as the transition probability from $|\psi\rangle$ to $|\phi\rangle$, or overlapping of them as follows:
\begin{equation}\label{eq:fidelity}
F(\rho,\sigma)=\left( Tr \left(\sqrt{\sqrt{\rho}\sigma\sqrt{\rho}} \right) \right)^2 = | \langle \psi | \phi \rangle   |^2 .
\end{equation}
Quantum errors can be regarded as probabilistic and discrete errors \cite{Devitt2013}  because they can be modeled by a set of error channels such as Pauli matrices. In this model, error probability $p$ is small enough so that the fidelity becomes
\begin{equation}\label{eq:error}
F(\rho,\sum E_k(p) \rho E^{\dagger}_k(p)  )=1 - cp +O(p^2),
\end{equation}
where ${E_k}$ is a set of error channels. Assuming $c=1$, the error probability directly affects the closeness of quantum states.

We assumed an error probability per unit time as $P_e$. A quantum gate operation gate error for time $T_G$ is defined as $P_e T_G$. A dephasing error probability for memory time $T_M$ is defined as $P_e T_M$. Therefore, the fidelity of physical qubit, $F_p$, is defined as
\begin{equation}\label{eq:PhyF}
F_p=1-P_M .
\end{equation}
The fidelity of logical qubit is defined as
\begin{equation}\label{eq:LogF}
\begin{split}
F_L=1-[1-\sum\limits_{k=0}^{1} {{n}\choose{k}}{P_M}^k (1-{P_M})^{n-k}]
\\-[1-(1-{P_{QEC}})^{n} ] +\left( \frac{P_M}{n} \right)\left( \frac{P_{QEC}}{n} \right),
\end{split}
\end{equation}
where $n$ is the number of qubits for QEC, and $P_M,P_{QEC}$ are the errors caused in memory time and QEC time, respectively.
The fidelity of logical qubit has four terms. The first term is unity because the final state is the same as the initial state if no error happens. The second term means occurring errors during memory time. 3-qubit QEC codes or 5-qubit QEC code can correct up to one error so that more than one error will deteriorate the original state. Hence, the term becomes the completion set of allowable error probability. The third term considers introduced errors during QEC time. The QEC codes cannot preserve the original data if an error occurs other than memory time. For that reason, QEC time, $T_{QEC}$ is defined as the sum of encoding, detection and correction times. The last term takes into account the probability that the two errors in the data qubit occurring in QEC and memory time will cancel each other.

The fidelity of logical qubit encoded with 5-qubit QEC codes, based on Eq. \ref{eq:LogF}, are deduced as follows:
\begin{equation}\label{eq:5QLogF}
\begin{split}
F_L=1-[1-5{P_M} (1-{P_M})^{4} - (1-{P_M})^{5}]
\\-[1-(1-{P_{QEC}})^{5} ] +\left( \frac{P_M P_{QEC}}{25} \right) .
\end{split}
\end{equation}
The fidelity gain is described by the fidelity difference between the physical and logical qubit,
\begin{equation}\label{eq:gain}
\begin{split}
G=F_L-F_P.
\end{split}
\end{equation}
A positive gain means that the accuracy of logical qubit is improved at the same computing time. This model is valid until the error probability is small. Nevertheless, the advantage of it is that it provides a rough range of computing time which is beneficial to the logical qubit. Thus, it can estimate the critical points of QEC codes such as $T^{memory}_{min}$, $T^{memory}_{max}$, and $T^{memory}_{best}$.

First, the minimum memory time is deduced at $G=0$. Since the minimum condition implies that the error probability is small, the logical qubit fidelity is approximated as $1-5P_{QEC}$. Therefore, the minimum memory time becomes
\begin{equation}\label{eq:minT}
T^{memory}_{min}=5T_{QEC} .
\end{equation}
The condition for deducing the maximum memory time is $G=0$ with $T_{QEC} \ll T_{memory}$. Therefore, the the equation becomes
$(1-P_M)\{(1-P_M)^3 (4P_M+1) -1)\}$. The solution of the polynomial equation of which domain is between $5P_{QEC}$ and 1 is
\begin{equation}\label{eq:maxT}
T^{memory}_{max}\cong \frac {0.1311} {P_e}.
\end{equation}
The best memory time is obtained at $G'(T_M)=0$. However, it is too difficult to solve fourth order polynomial equation with an arbitrary constant $T_{QEC}$. Employing $T_{QEC}$ is short enough, $G'(T_M,T_{QEC})= G''(\alpha)(T_M-\alpha)+G'(\alpha)$, where $G'(\alpha,0)=0.$ Then, the best memory time becomes
\begin{equation}\label{eq:bestT}
T^{memory}_{best}\cong \frac {0.06-0.003P_{QEC}} {P_e} .
\end{equation}

This model estimate that there is a rough range of memory time, $5T_{QEC} < T_{memory} <  {0.1311}/ {P_e}$, to show that the quantum gate operation with logical qubit is accurate. In addition, the gain will be maximum at Eq. \ref{eq:bestT}.
The whole time for QEC procedure is, therefore,
\begin{eqnarray}\label{eq:TQECcycle}
T_{QEC cycle}=T_{QEC} +T_{memory,best} \nonumber\\
\cong \frac {0.06+0.997P_{QEC}} {P_e}.
\end{eqnarray}
In this regard, given the coherence time and gate error rate, optimal QEC cycle exist. Considering a longer coherence time, the error rate $P_e$ is decreased, and therefore a longer computing time is obtained.

\section{\label{sec:level1} DENSITY-MATRIX SIMULATION}

The density-matrix simulation has several advantages to describe the behavior of qubits.
The amount of occurring error depends on the quantum states and error channels.
In spite of the same error probability $p$ for X- and Z-type errors, this simulation copes with the error differently.
For example, if the initial state is $|1\rangle$, X error lower the fidelity but Z error does not. Two qubit gates are also described in a different way. Suppose that X-type error is introduced on the controlled qubit of a CNOT gate. The error not only change the controlled qubit but also changes the state of target qubit after the CNOT gate.
In addition, those errors are described in physical quantities such as $T_1$ relaxation, $T_2$ dephasing time. Hence, the error channels are expected to describe the quantum states more close to the actual quantum states. We focus on several specific states to find the gain of fidelity without calculating the minimum fidelity gain of every possible states. This is because examining an opportunity to take an advantage of QEC codes with physical qubits in specific states is a ongoing problem in small qubit systems.

\begin{figure}
\includegraphics[width=\linewidth]{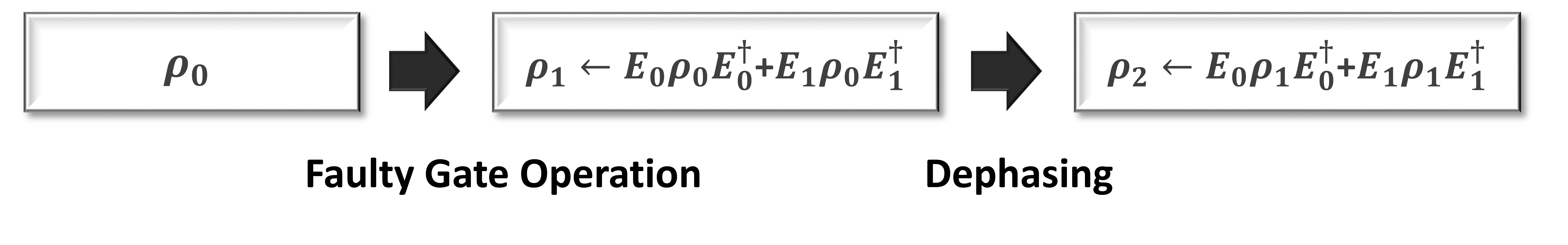}
\caption{\label{fig:error} Evolution procedure for operating quantum gates and introducing errors. First step is to conduct faulty quantum gate operation consisting of successful gate operation $E_0$ and gate error $E_1$. The last step introduces relaxation and dephasing error, sequentially. During the gate operation time, faulty gate operation is conducted immediately and single qubit errors are made.}
\end{figure}

The evolution of density matrix takes two steps: faulty gate operation and independent qubit errors. The error are described as a complete positive mapping. In faulty gate operations, perfect operation $E_0$ occurs with the probability $1-p$, and imperfect operation $E_1$ does with probability $p$ as follows:
\begin{eqnarray}\label{eq:faultygate}
E_{0}=\sqrt{1-p} U_{gate} ,\,\, E_{1}= \sqrt{p} A U_{gate} = \sqrt{p} I,
\end{eqnarray}
where $U \in \{ X,Y,Z,CNOT,SWAP\}$.
The transformation process is described in Fig. \ref{fig:error}. The operation $A$ is properly chosen by gate operation $U_{gate}$ and it becomes identity for the gates in a set $U$.
The decoherence error during a gate operation is dealt with separately from the gate operation error. This is because the gate operation time compared to coherence time is so small that it can be done shortly and then decoherence errors follow during the gate operation time.

The decoherence errors consists of $T_1$ relaxation, $T_2$ dephasing errors. Both errors are described in amplitude damping and phase error models. Error probabilities in these models are derived from the $T_1$ and $T_2$ coherence time so that the evolution of quantum state is likely to express practical circumstances.
The amplitude damping channel is described as follows:
\begin{eqnarray}\label{eq:dephasing1}
E_{0}^{amp}=
\begin{pmatrix}
1  &\sqrt{\gamma}  \\
0  &1  \\
\end{pmatrix}
, E_{1}^{amp}=
\begin{pmatrix}
1  &0  \\
0  &\sqrt{1-\gamma}  \\
\end{pmatrix},
\end{eqnarray}
where  $\gamma = 1- e^{t/T_1}$.
The error probability $\gamma$ is derived from the $T_1$ relaxation time. Similarly, the phase error model is described as follows:
\begin{eqnarray}\label{eq:dephasing2}
E_{0}^{phase}=
\sqrt{\alpha} \begin{pmatrix}
1  &0  \\
0  &1  \\
\end{pmatrix}
, E_{1}^{phase}=
\sqrt{1-\alpha}\begin{pmatrix}
1  &0  \\
0  &-1  \\
\end{pmatrix},
\end{eqnarray}
where  $\alpha = (1+e^{-t/2T_2})/2$.
The error probability $1-\alpha$ is originated from the $T_2$ dephasing time.
In idle time and after the gate operation, relaxation errors are followed by dephasing errors. Assuming that the initialization and measurement processes are perfect, we focused on the gate error and decoherence errors.

\section{\label{sec:level1} Analysis}

In this section, we present density-matrix simulation results when the use of logical qubit is beneficial according to the memory time. The gain is defined as the difference of fidelities between a logical qubit and a physical qubit described in Eq. \ref{eq:gain}. Coherence time, $T_{c}$, is assumed to be 1 as unit time not for losing generality of simulation. A controlled-Z gate is used as the primitive 2-qubit gate with nearest neighbor interaction, and according gate operation time is used as the ratio between the actual operation time and coherence time \cite{Veldhorst2015}. Qubits are subjected to individual dephasing processes and the gate operation error.

\subsection{\label{sec:level2} Comparison with Linear Approximation Model}

\begin{figure} 
\includegraphics[width=\linewidth]{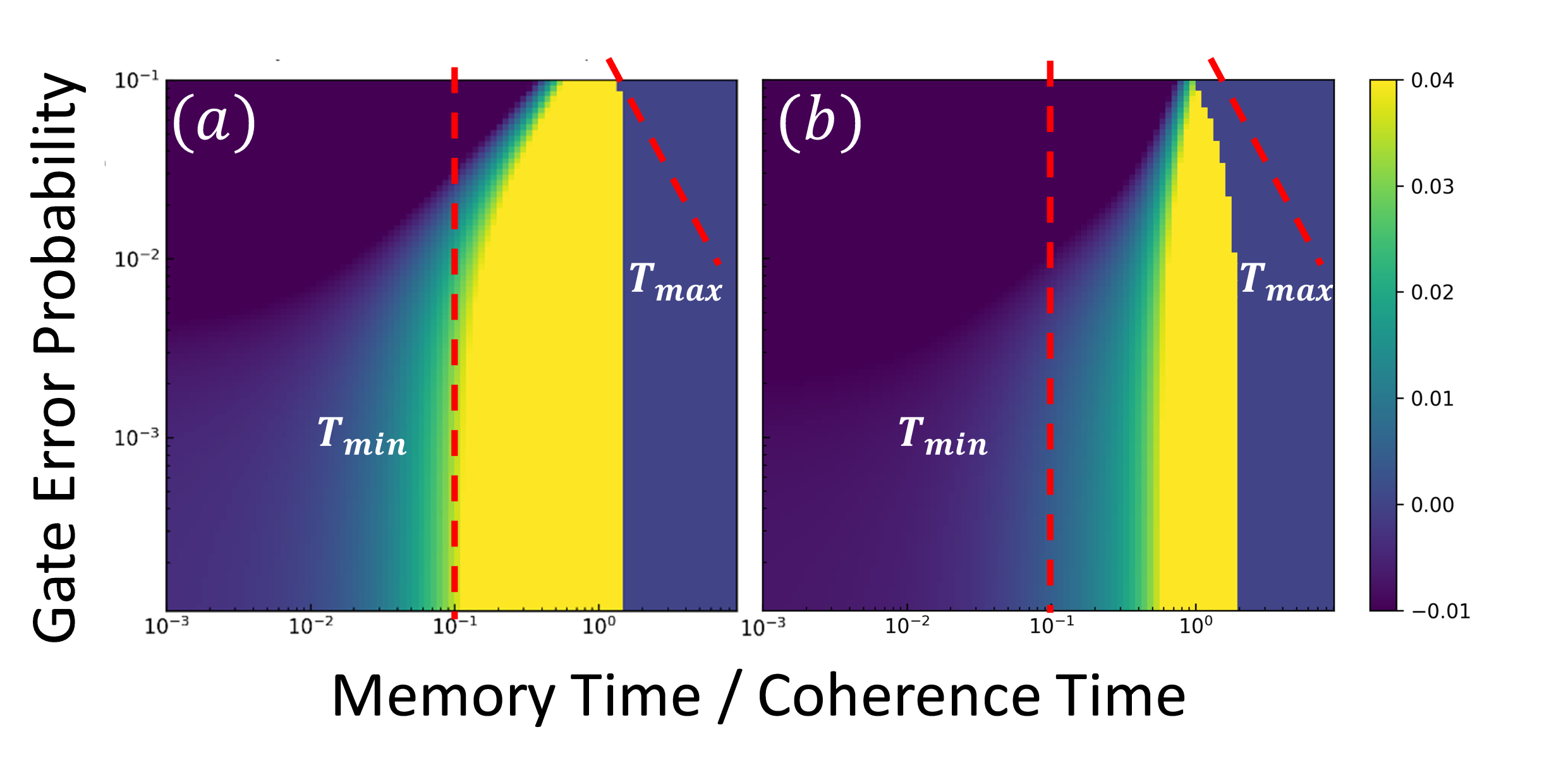}
\caption{\label{fig:consistency}  Comparison of linear approximation model and density matrix model for 5-qubit QEC code. (a) When the coherence time is considered as 1$ s $ and the QEC time is 0.02 $ T_c $ with the initial $|1\rangle$ state, the estimates of the two models are parallel. (b) In the initial $R_x(\frac{\pi}{4})|1\rangle$  state with QEC time of 0.018$\times T_{c}$, the estimation from linear approximation is unnecessarily wide so that it can be used as a minimum condition for using logical qubits}
\end{figure}

To achieve a positive gain for using logical qubits, both linear approximation and density matrix models predict the beneficial memory time depending on gate error probability. The first model provides a direct relation between the fidelity of the logical qubit and error probabilities so that the range of memory time is calculated. On the other hand, density-matrix calculation take into account the detailed procedures such as the initial state, type of QEC codes, and the layout of circuits so that a feasible results is expected.

The two models have a common point to estimate the requirement when logical qubit is beneficial. Considering the logical $|1\rangle$ state encoded by 5-qubit QEC codes with QEC time of 0.02 $T_c$, the estimated minimum and maximum memory time for linear approximation model are consistent with those for density-matrix model, which is shown in the Fig. \ref{fig:consistency}(a). However, the expectation based on linear approximation is not always valid. For instance, it expects the same fidelity for the initial state is $|1\rangle$ or $|+\rangle$ state. This is because the gate error probability still remains the same. Nonetheless, $|1\rangle$ is deteriorated by $T_1$ dephasing, while  $|+\rangle$ state goes through both $T_1$ and $T_2$ dephasing. Moreover, the effect of gate time on the fidelity is different. If the gate operation time is shortened with the same error probability, the estimation of minimum memory time is much smaller than that based on the density-matrix model, which is illustrated in Fig. \ref{fig:consistency}(b). In this regard, the expected range based on the linear approximation is useful to have intuition for a necessary condition.

\subsection{\label{sec:level2} IBM QX Results}

\begin{figure}
\includegraphics[width=\linewidth]{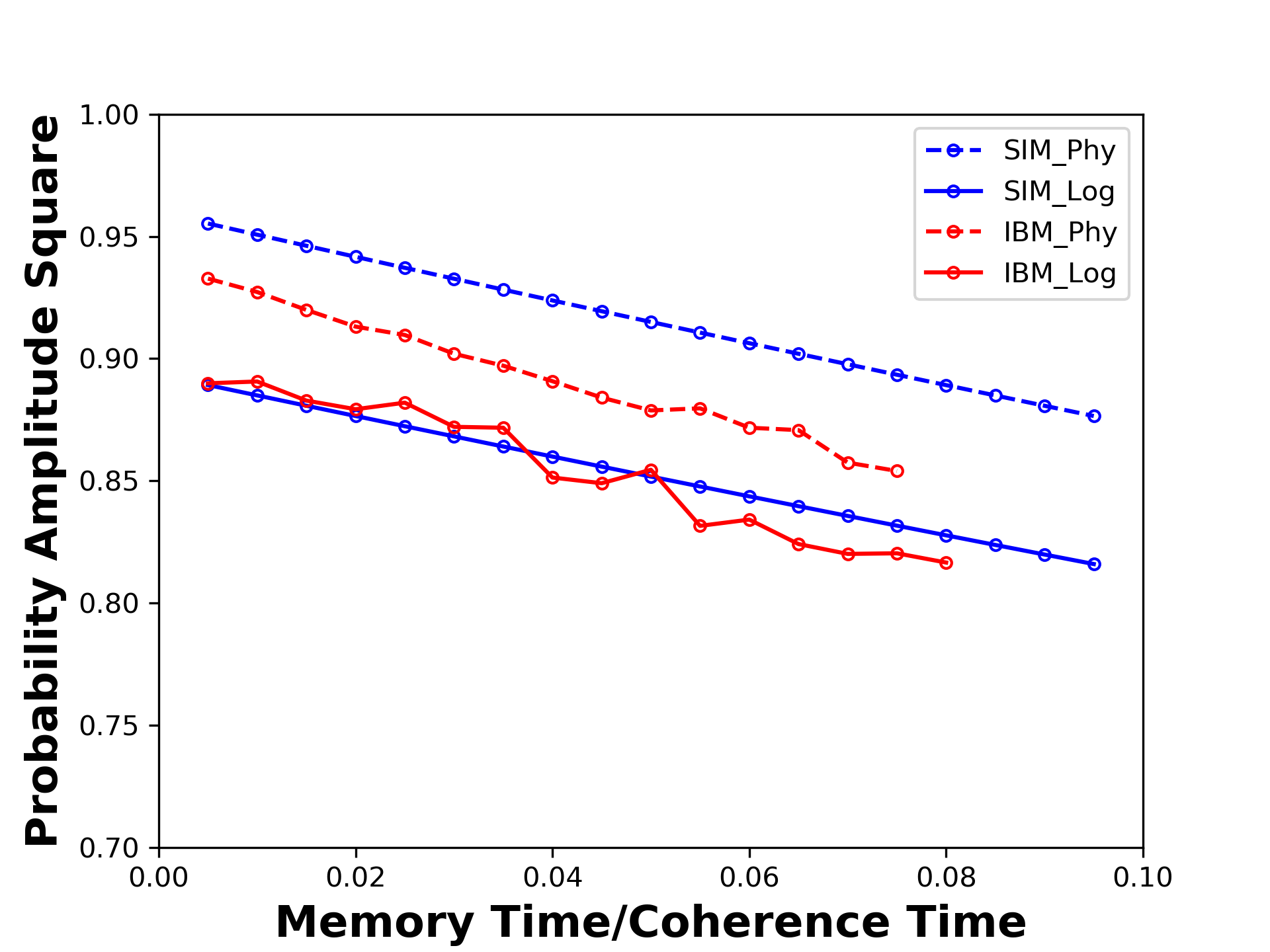}
\caption{\label{fig:IBM} Comparison of density-matrix simulation(blue lines) and IBM quantum experience result(red lines). The probability of spin up state is described according to the memory time. The spin up state probability of physical qubits are dashed lines, while that of logical qubits are solid lines. Simulation result expects the probability of a logical qubit is not higher than that of a physical qubit with the same memory time. However, the agreement of simulation and experimental result shows that the density-matrix simulation is trustworthy.
}
\end{figure}

IBM quantum experience(IBM QX) provides 5 superconducting qubit system which is available via IBM cloud platform. The average $T_1$ and $T_2$ coherence times of qubits are 54 $\mu s$ and 44 $\mu s$, respectively. The 1-qubit gate operation is composed of X and Z gates \cite{McKay2017}. To prevent abrupt change, the time interval between quantum gates are set as 10 $ns$. The accuracy of 1-qubit gate operation is over 99.9\%, and the average accuracy of 2-qubit gate operation is 99\%. The operation times of H and CNOT gate are 50 $ns$ and 122 $ns$, respectively.

In this density-matrix simulation, the initialization and measurement errors are taken into account. From various QEC codes, the 3-qubit QEC circuit, descried in Fig \ref{fig:circuit}, is chosen due to the simplicity. In Fig \ref{fig:IBM}, the blue and red dashed lines are the probability amplitude square of physical qubits in simulation and IBM QX, respectively. The probability is obtained from the results of 8092 repetitions. The blue and red solid lines are that for logical qubits in simulation and IBM QX, respectively. There is no cross point between a physical and logical qubit. It means that given 3-qubit QEC and qubit properties, the logical qubit is not accurate in spite of the length of memory time. However, this result does not deny the accuracy of simulation. Rather, the result is concurrent with the simulation. Based on the consistency, we propose that the performance of logical qubits can be predicted in the simulation.

\section{\label{sec:level1} Application}
\subsection{\label{sec:level2} QEC Dependency}
\begin{figure}
\includegraphics[width=\linewidth]{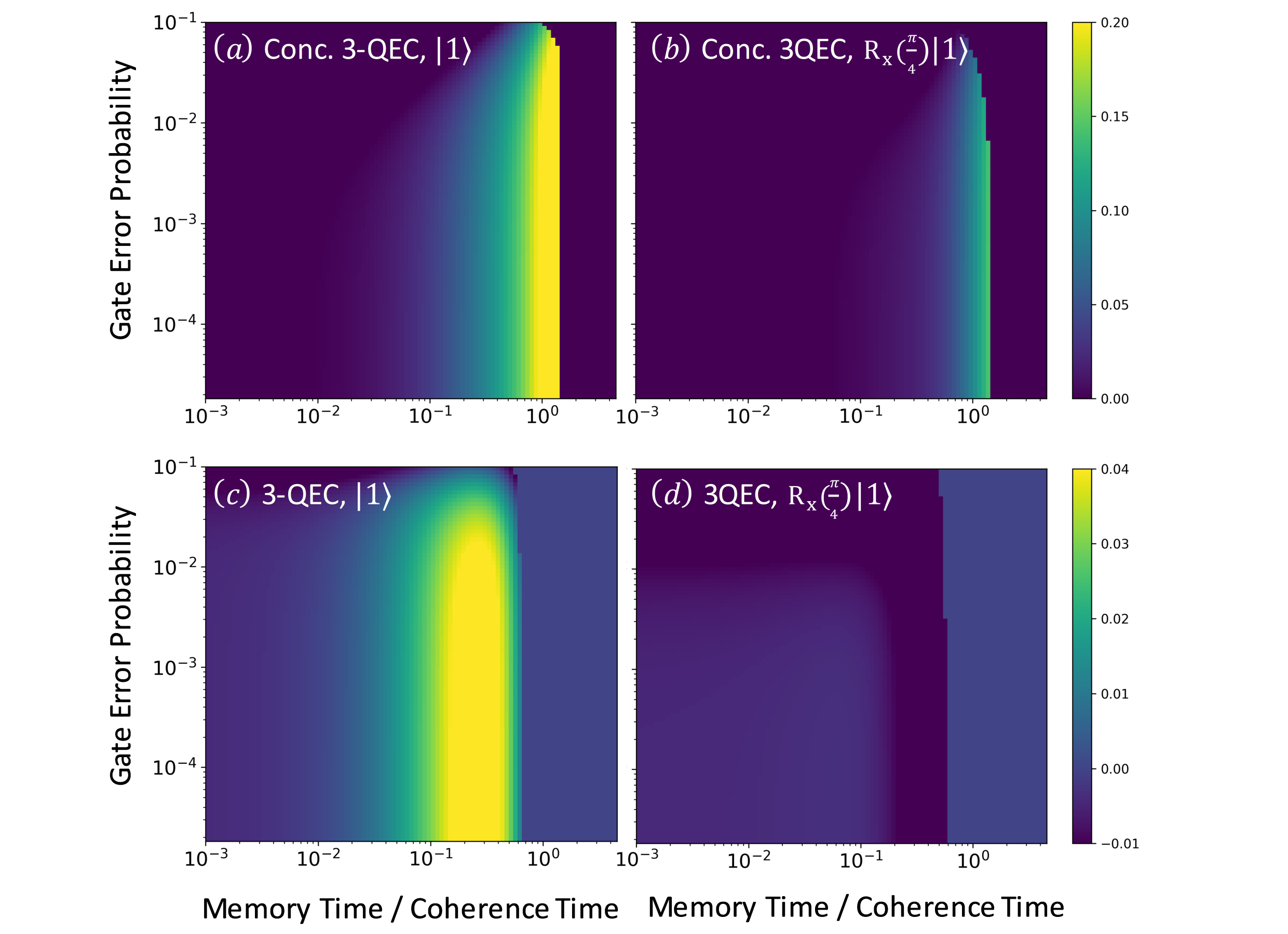}
\caption{\label{fig:state} (a-b) 5-qubit QEC code with the initial state is $|1\rangle$ state, and $R_x(\frac{\pi}{4})|1\rangle$ states, respectively. The maximum gain is 0.25 and 0.13, respectively. (c-d) A bit flip 3-qubit QEC code with the initial state is $|1\rangle$ state, and $R_x(\frac{\pi}{4})|1\rangle$ states, respectively. The maximum positive gain is 0.05 for $|1\rangle$, while there is no positive gain for $R_x(\frac{\pi}{4})|1\rangle$. The bit flip 3-QEC code cannot correct phase errors.
}
\end{figure}
The advantage of density-matrix simulation is that the gain of logical qubit can be evaluated according to initial states. For example, $|1\rangle$ state only suffers from $T_1$ dephasing, while $|+\rangle$ does from $T_1$ and $T_2$ dephasing. Unlike a bit-flip 3-qubit QEC code, 5-qubit QEC codes are supposed to correct both X- and Z-type errors. Therefore, it is expected to figure out the differences in different QEC codes.

The simulation result for a logical qubit encoded by 5-qubit QEC is shown in Fig. \ref{fig:state}(a)-(b). The negative or zero gain is obtained when the memory time is too short. This is because the fidelity of logical qubit is reflected in the gate error and dephasing error during the encoding, decoding and correction process. If the memory time is too long, both physical and logical qubit lose their quantum information, and the gain goes to zero. To prevent the gain from being exaggerated, it is set to 0 when the fidelity of logical qubit is below 0.75 and that of physical qubit is below 0.5. In the intermediate region, the competition of benefits and costs determines the gain of logical qubits.

For the initial state $|1\rangle$, the range for obtaining the positive gain is increased as the gate error probability decreases, which is shown in Fig. \ref{fig:state}(a). If the error probability is 0.1\%, the memory time should be between 0.45-1.1$T_{c}$, and in that range, its ,maximum gain is 0.13. If the error probability is 0.01\%, the maximum gain is increased up to 0.25. Compared to the 3-qubit QEC code, the maximum gain is merely 0.05, which is shown in \ref{fig:state}(c), even though the operation time and gate error rate is the same.

The gate operation time is 0.001$T_{c}$. QEC time for 5-qubit QEC codes is 0.018$T_{c}$, while QEC time for bit flip 3-qubit QEC codes is 0.010$T_{c}$. The $x$ axis is memory time per coherence time so that this results can be applied into other systems such as superconducting system. The $y$ axis is gate error probability for both 1-qubit and 2-qubit gates.

5-qubit QEC also shows better performance for states including phase errors. The initial state of $R_x(\frac{\pi}{4})|1\rangle$ is affected by both bit and phase flip errors. If a logical state is encoded by 5-qubit QEC, the maximum gain is up to 0.14, which is illustrated in \ref{fig:state}(b). On the other hand, bit flip 3-qubit QEC, which is not expected to correct phase errors, does not have any positive gain, described in \ref{fig:state}(d). Therefore, the 5-qubit QEC is able to correct both bit and phase errors, and higher fidelity is obtained.

\subsection{\label{sec:level2} Layout Dependency}
\begin{figure}
\includegraphics[width=\linewidth]{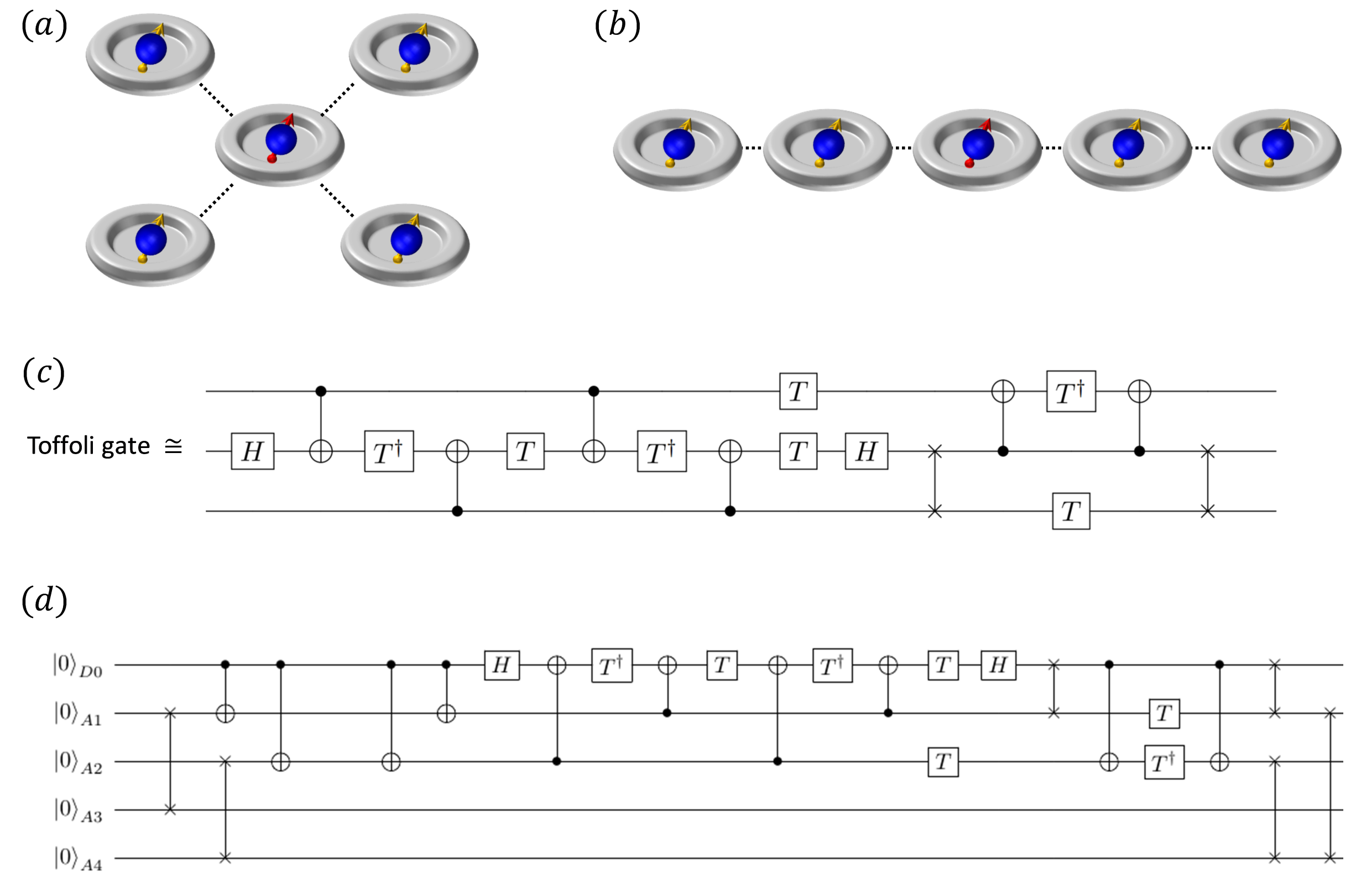}
\caption{\label{fig:circuitlayout} (a-b) 5-qubit layouts. The red arrow qubit in the middle contains data, while the other yellow qubits are ancilla. The 2-qubit nearest neighbor interaction between qubits are depicted in dotted line. (c) Decomposed circuit for a toffoli gate. Two ancilla is placed at the ends and data qubits are in the center with nearest neighbor interaction. (d) The 5-qubit QEC circuit with nearest neighbor interaction. The level-1 logical qubit is encoded by 3-qubit phase flip QEC code. The level-2 logical qubit is encoded with 3-qubit bit flip QEC code}
\end{figure}

In the current technology, it is difficult to establish qubits which are all connected with each other. The connection problem becomes important not only for multi-qubit systems but for 3-qubit QEC codes. This is because partial connection among qubits demands additional SWAP gates. In this work, we used a controlled-Z gates as a primitive operation, which is already implemented in quantum dot qubit system. A Toffoli gate is decomposed into 6 controlled-Z gates and single qubit gates. The controlled-Z gate is advantageous to implement CNOT gate with both directions because the same number of 1-qubit operations is required.

The 5-qubit QEC codes are illustrated in Fig. \ref{fig:circuit}(b). We take into account two layouts: X layout and linear layout, which are described in Fig. \ref {fig:circuitlayout}(a) and \ref {fig:circuitlayout}(b), respectively. These two layouts are a promising candidate for fabricating qubits \cite{ibmqx2,Ito2016} . Considering nearest neighbor interaction, data qubit should be placed at the center of both layouts so that the minimum number of SWAP gates is added. Although X layout is close to all-to-all connection, a decomposed toffoli gate asks for two SWAP gates, shown in Fig. \ref {fig:circuitlayout}(c). The linear layout is shown in Fig. \ref{fig:circuitlayout}(d). Qubit A1 and A2, located next to the data qubit, preserve phase flip error in the data qubit. Qubit A3 and A4, placed at the both ends, does bit flip error in data qubit. Comparing two layouts to perform 5-qubit QEC codes, the X layout needs additional 2 SWAP gates, while linear layout does 6 SWAP gates.

\begin{figure} 
\includegraphics[width=\linewidth]{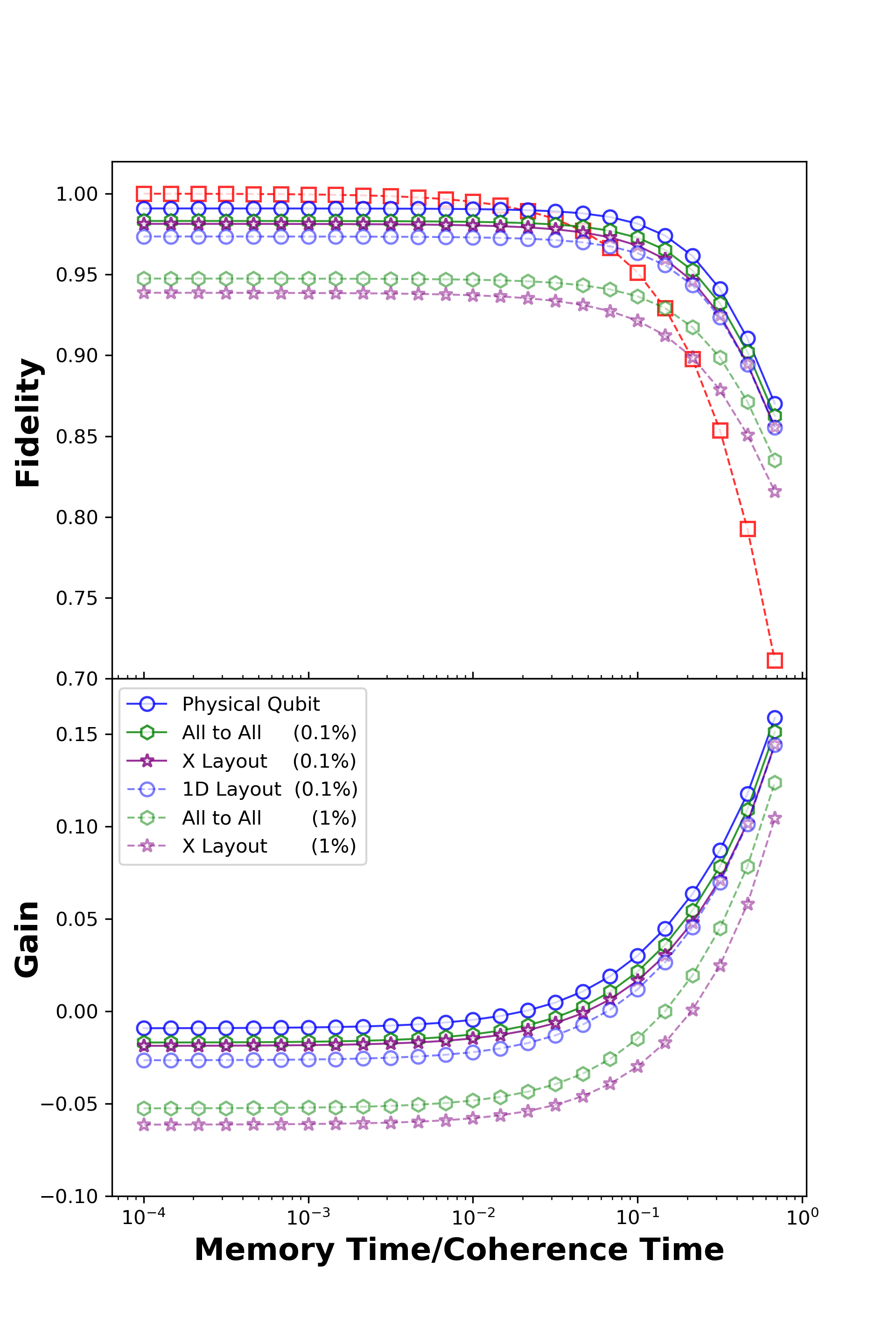}
\caption{\label{fig:layout} The same 5-qubit QEC circuit with different layouts. (a) In all-to all connection, no additional SWAP operations are required. (b) In X layout, 2 SWAP gates are added. (c) In linear layout, 6 more SWAP gates are necessary. Since a single SWAP gate is decomposed to 3 controlled-Z gates and 6 hadamard gates, the gate errors and dephasing errors during the operations are included. To achieve the positive gain for using logical qubits, both models predict the beneficial memory time depending on gate error probability. }
\end{figure}

Considering nearest neighbor interaction, the qubit layout determines the number of additional SWAP gates. The overhead cost for conducting a given quantum circuit becomes important as the number of qubits and operations increases. To perform the 5-qubit QEC code, X and linear layout needs 2 and 6 more SWAP gates. Since a SWAP gate is decomposed into 3 CNOT gates, a SWAP gate required 3 controlled-Z gates and 1-qubit gates in primitive operations for quantum dots. The number of additional gates is much larger to implement a large-scale quantum circuit. Even though the accuracy for 2-qubit gate operation is still improving, it is difficult to implement all-to-all connected circuit. Hence, it is worth knowing the impact of the overhead cost on the fidelity.

This simulation is designed to weigh whether the qubit layout is important or gate accuracy is important. The operation time and accuracy of controlled-Z gates are changed to calculate the fidelity of qubits, as shown in Fig. \ref{fig:layout}. As the operation time increases, more dephasing errors are accumulated. The error probability has a direct effect on the performance of a circuit. Considering the 99.9\% accuracy of gate operation, the all-to-all connected layout have the maximum gain of 0.4. The X and 1D linear layout have the maximum gain of 0.3 and 0.2, respectively. If the gate accuracy is lowered to 99\%, the discrepancy between the all-to-all connected layout and the other layouts is increased. Consequently, the miscalculation by additional gates is increased.

\begin{figure}
\includegraphics[width=\linewidth]{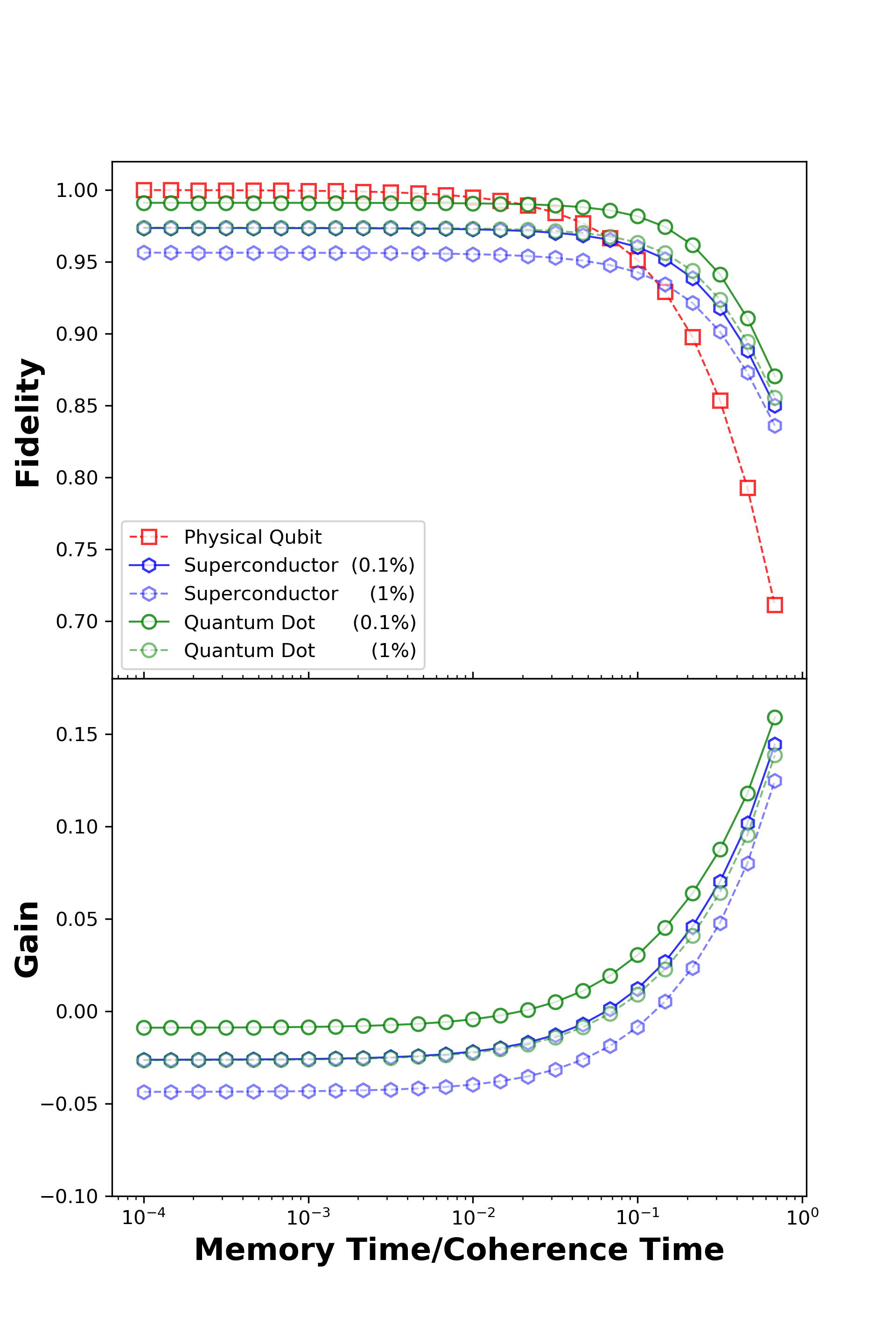}
\caption{\label{fig:device} The expected performance evaluation of physical qubits and logical qubits of superconducting qubits and quantum dot qubits. Logical qubits are encoded by 5-qubit QEC codes. The initial state is $|1\rangle$ state. The reported gate operation time is used\cite{Barends2014, Kelly2015,Yoneda2018,Veldhorst2015}. The fidelities of logical qubit composed of superconducting qubits of 1\% gate error and 0.1\% gate error are blue hexagon line and light blue hexagon line, respectively. Those composed of quantum dot qubits are green circle line and light green circle line, respectively.}
\end{figure}

\subsection{\label{sec:level2} Evaluations of Physical Qubits}
\begin{table} [b] 
\caption{\label{tab:table1}%
The reported properties of superconducting and quantum dot qubits: coherence time $T_{c}$, 1-qubit gate operation time $T_{1Q}$, and 2-qubit gate operation time $T_{2Q}$ is summarized.}
\begin{ruledtabular}
\begin{tabular}{crr}
\textrm{Value}&
\textrm{Superconductor} \cite{Barends2014, Kelly2015}&
\textrm{Quantum Dot} \cite{Yoneda2018,Veldhorst2015}\\
\colrule
$T_{c}$ & 10 $\mu s$ & 120 $\mu s$\\
$T_{1Q}$ & 20 $ns$ & 130 $ns$\\
$T_{2Q}$ & 40 $ns$ & 100 $ns$\\
\end{tabular}
\end{ruledtabular}
\end{table}

The performance of logical qubits is directly influenced by the gate operation time and coherence time. The number of operation is often considered as the ratio between them. To evaluate the expected performance of logical qubits encoded by 5-qubit QEC codes, recent superconducting qubits \cite{Barends2014, Kelly2015} and quantum dot qubits \cite{Yoneda2018,Veldhorst2015} are used. The data is summarized in Table \ref{tab:table1}. Even though 5 qubits in quantum dot system have not been fabricated yet, it is assumed that hypothetical 5 qubits are placed in the layouts.

In the simulation, the reported operation time and coherence time is used. $T_1$, and $T^*_2$ is used as $T_{c}$ in superconducting and quantum dot qubits, respectively. All of the 1-qubit gates are assumed to have the same operation time, and a 2-qubit gate of controlled-Z operation is implemented. The accuracy of gate operation is considered as the same for 1-qubit and 2-qubit operation as 99\% or 99.9\%. For the gate accuracy of 99.9\%, logical superconducting qubit and physical qubit have the same fidelity of 0.96 at time 0.13$T_{c}$. In quantum dot qubit, given the same gate accuracy, the fidelity of logical qubit is higher than that of physical qubit if the calculation time is longer than 0.07$T_{c}$, which is illustrated in Fig. \ref{fig:device}. If the accuracy of gate operation is increased to 99.9\%, the intersection point of the fidelity of logical and physical qubits becomes 0.07$T_{c}$ and 0.025$T_{c}$ and the maximum gain becomes 0.14 and 0.16 for superconducting and quantum dot qubit, respectively.
It implies that the number of possible gate operations is increased, assuming that the connectivity of the qubits and additional errors are ignored and that the logical gates are traversal. For example, at the same 90\% fidelity, physical qubits can conduct 230 gates, while logical qubits can do 330-520 gates.

\section{\label{sec:level1}CONCLUSION}
In this paper, we suggest three application examples using density-matrix simulation. First, the performance of logical qubits encoded by arbitrary QEC codes can be estimated.
Second, the importance of layouts can be estimated, given the gate error probability.
Third, the performance of logical qubits can be estimated based on the data of the current physical qubits, such as coherence time, operation time, and gate accuracy.
investigate how to obtain the longest computing time in density-matrix simulation, using 3-qubit QEC codes. 3-qubit QEC codes are so simple that the additional SWAP gates are minimal despite nearest neighbor interaction. The primitive gate sets and gate operation time are referred to the data for spin quantum dot qubits. Given the error probability of 0.1\% in the all-to-all connected layout, the gain for using a logical qubit is up to 0.25 for $|1\rangle$.

In density-matrix model, the simulation results takes into account precise error channel for bit and phase errors. The error probabilities are reflected in terms of $T_1$ and $T_2$ coherence time. The exhaustive simulation results are conducted to figure out the dependency on the initial states, QEC codes, and qubit layouts.
The linear approximation can estimate the minimum and maximum condition of memory time expressed in gate errors and QEC time to obtain positive gain. Nevertheless, it is regarded as the minimum requirement because error types and error propagation is neglected. In this work, we scrutinize the existence and importance of the beneficial condition for logical qubits. It is expected to be a proper milestone for implementing a logical qubit to operate in a longer computing time.

\begin{acknowledgments}
This work was supported by Electronics and Telecommunications Research Institute (ETRI) grant funded by the Korean government. [18ZH1400, Research and Development of Quantum Computing Platform and its Cost-Effectiveness Improvement]

The ImPACT Program of Council for Science, Technology and Innovation (Cabinet Office, Government of Japan) the Grant-in-Aid for Scientific Research (No. 26220710, 16H00817, 17H05187) CREST (JPMJCR15N2, JPMJCR1675), PRESTO (JPMJPR16N3), JST Incentive Research Project from RIKEN The Telecommunications Advancement Foundation Research Grant Futaba Electronics Memorial Foundation Research Grant Kato Foundation for Promotion of Science Research Grant, Hitachi Global Foundation Kurata Grant,
The Okawa Foundation for Information and Telecommunications Research Grant, The Nakajima Foundation Research Grant, Japan Prize Foundation Research Grant

\end{acknowledgments}


\providecommand{\noopsort}[1]{}\providecommand{\singleletter}[1]{#1}%

\end{document}